# Design and properties of low energy x-ray transmission windows based on graphenic carbon

**Sebastian Huebner**[*,1], **Natsuki Miyakawa**[2], **Andreas Pahlke**[2], **and Franz Kreupl**[1]

[1] Department of Hybrid Electronic Systems, Technische Universität München, Arcisstr. 21, 80333 Munich, Germany
[2] Ketek GmbH, Hofer Str. 3, 81737 Munich, Germany

X-ray transmission windows for the low energy range, especially between 0.1 keV and 1 keV have been designed and fabricated based on graphenic carbon (GC) with an integrated silicon frame. A hexagonal and a bar grid support structure design have been evaluated. The bar grid design allows to substitute polymer-based windows with the advantages of higher transmission, better rejection of visible light and vacuum operability of the encapsulated silicon drift detectors (SDD). In addition, the high mechanical resilience of graphenic carbon is demonstrated by pressure cycle tests, yielding over 10 million cycles without damage. The data are complemented by bulge tests to determine a Young`s modulus for graphenic carbon of approximately 130 GPa. Additional finite-element simulation and Raman studies reveal that the mechanical stress is not homogeneously distributed, but reaches a maximum near the anchoring points of the free standing graphenic carbon membrane.

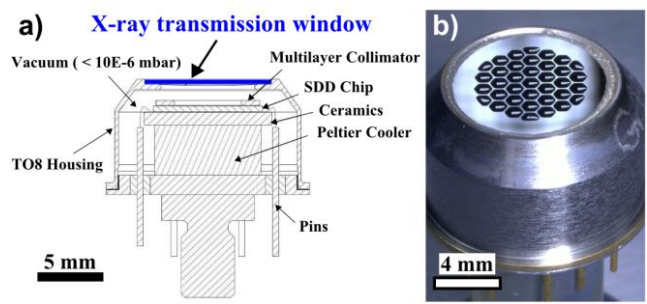

The cross section of a vacuum encapsulated silicon drift detector (SDD) in (a) shows the highly transparent x-ray transmission window that is integrated into the top of a TO8 housing. (b) shows a prototype TO8 detector module with a graphenic carbon, low energy transmission window with an hexagonal support grid structure.

**1 Introduction** Silicon drift detectors (SDD) are used for energy dispersive spectroscopy (EDS) and x-ray fluorescents (XRF) applications [1-3]. Proper and reliable operation of SDDs requires a hermetically tight housing with an integrated, highly x-ray transparent window, shown in Fig. 1. Placing the detector under a high vacuum allows for efficient detector cooling and avoids the contamination of the detector surface. The resulting pressure load of one atmosphere weighs on the x-ray window, which is a significant mechanical burden for the window material.

Traditionally, transmission windows are made out of beryllium (Be) due to the high x-ray transparency and mechanical strength of Be that allows the fabrication of transmission windows that do not require a support structure [4]. However, for a 7 mm wide, circular unsupported window structure, the Be windows need to be several micrometres thick in order to provide a gas tight configuration [4]. This leads to a high attenuation of low energy x-ray radiation in the range below 1 keV. Therefore, light element detection in EDS applications is not possible with Be transmission windows [4]. Figure 1c shows the relevant characteristic $K_\alpha$ x-ray line energies of elements that are found in the given energy range and compares it to the transmission of an 8 μm thick Be-window. Several important elements including boron (183 eV), carbon

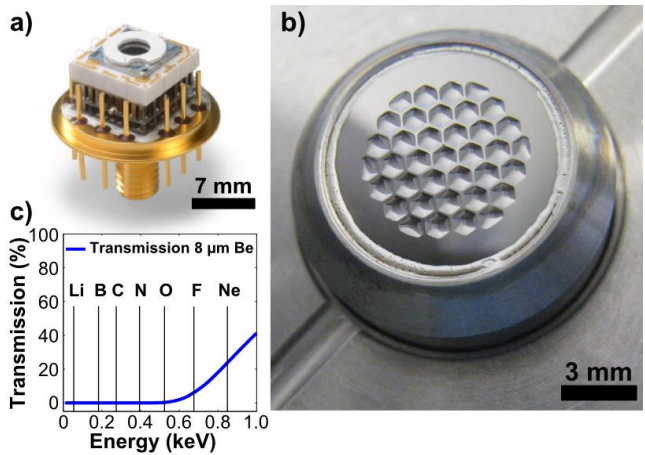

**Figure 1** A SDD detector without TO8 housing is shown in (a). The gas tight TO8 housing with a highly transparent x-ray transmission window based on GC, shown in (b), will hermetically seal the SDD. Figure (c) shows the x-rays transparency of a Be window and the energies of the characteristic K$\alpha$ energy lines of elements in the low energy range [5, 6].

(277 eV), nitrogen (392 eV), and oxygen (525 eV) can therefore not be detected using Be-windows [5, 6].

As low energy x-ray radiation is readily absorbed by most materials, a suitable low energy transmission window needs to be extremely thin which is contradictory to the necessary high mechanical stability. The mechanical requirements of low energy transmission windows depend on the desired window diameter and the differential pressure that acts on the transmission window. One way to ease the stability requirements is by the implementation of a support grid in order to divide the open window geometry in numerous window elements with a reduced span width.

Such a support structure reduces the open transmission area and covers a significant amount of the window geometry, which reduces the detector efficiency as the support structure is opaque for low energy radiation. This signal loss is quantified by a corresponding fill factor that relates the open area of the window to the total area of the window [7].

Polymer windows are the most common choice for low energy transmission windows and achieve an acceptable x-ray transmission for radiation in the energy range of 0.1 keV – 2 keV by using low Z elements and a reduced window thickness [7, 8]. The low temperature tolerance of polymer windows hinders vacuum encapsulation and limits the performance of detector modules that incorporate a polymer transmission window. The low mechanical strength of the polymer films only allows a very limited span width in order to support the required differential pressure of greater than 1 bar and makes the use of a support grid necessary. This leads to a fill factor below 80 % [7]. The high optical transparency of the polymer film requires an aluminium coating to reach light blocking levels suitable for EDS applications, which further reduces the x-ray transmission and results in a spectral contamination due to x-ray fluorescence of the light blocking layer [9].

More recently, silicon nitride has been successfully implemented as a window material for low energy x-ray transmission, overcoming some of the limitations of polymer windows. Silicon nitride low energy transmission windows offer superior transmission in the energy range of 0.1 keV – 2 keV, high gas tightness, high temperature tolerance, and high chemical stability [10]. An 8 µm thick polycrystalline silicon support grid results in a fill factor of 77 % and high mechanical strength [11]. Similar to polymer transmission windows, the low attenuation of silicon nitride in the optical spectrum requires a light blocking layer if the detector is not used in a completely dark environment, with the implications of reduced transmission and spectrum contamination [10].

On the other hand, the properties of graphene have made it a promising candidate for extremely thin x-ray transmission windows due to the high mechanical strength [12], the demonstrated gas tightness of a single mono layer [13], the high chemical stability [14], the high electrical conductivity [15], which avoids charging, and the low atomic number of carbon.

The fact that graphene, transferred or grown by the traditional synthesis methods [16-18], only adheres to the target substrate via van der Waals forces [19] leads to delamination of fabricated graphene membranes instead of material failure, if a large enough differential pressure is applied. This limits the use of graphene as a window material as the poor adhesion becomes even more critical for larger open geometries than those described by Koenig et al. [19].

It was recently demonstrated that graphenic carbon (GC) is a suitable window material [20] as it is deposited directly from a gas phase onto a silicon substrate and overcomes the inadequate adhesion by forming strong silicon carbon bonds [21]. The inherently high bonding of the GC window material to a silicon frame not only hinders delamination but also acts as a gas tight barrier in contrast to graphene that adheres by van der Waals forces [13]. The 1 µm thick GC windows outperform beryllium as a window material for x-ray transmission windows [20], offering increased x-ray transmission, high mechanical and chemical stability and gas tightness, while avoiding the health concerns of beryllium [22-24]. The GC windows have a low attenuation of x-ray radiation for energies above 1 keV and even enable the detection of the low energy K$_\alpha$ lines of carbon and fluorine, but have a poor transmission for the lithium, nitrogen and oxygen K$_\alpha$ lines [20]. Therefore, the development and properties of GC windows with a reduced thickness, that allows a sufficiently high transparency for the energy range of 0.1 keV to 2 keV are demonstrated and discussed in this paper.

As x-ray transmission windows are also used in harsh environments, the resistance against ionizing radiation and against ozone attack are examined, which were identified as potential threats to the window integrity.

While GC shows a high resistance against most chemical compounds, ozone is known to attack carbon compounds. This could pose a problem at ambient conditions and the impact of an ozone containing environment is therefore evaluated [25, 26].

Zhou et al. [27] showed that soft x-ray radiation is a threat to the structural integrity of graphene. Therefore, the influence of a high dose of incoming ionizing radiation on the GC window material is examined as well.

In order to give more insights into the properties of the GC window material, the Young's modulus, which is an important parameter describing the mechanical properties of a material, is estimated. The discussion of the obtained results is complemented by Raman measurements and results from finite-element analysis.

**2 Window Fabrication and Experimental Procedures** The GC x-ray transmission windows are fabricated by depositing the GC material onto a silicon substrate by a CVD process described in reference [20]. They are subsequently glued into the end cap of a TO8 housing for further investigation.

Two fabrication routes for low energy transmission windows are validated in this work. One approach is the deposition of the GC material onto bulk silicon substrates which are subsequently structured to form the support grid and the other, the deposition on a pre-structured silicon substrate. The former has the advantage of a simple fabrication process as the GC material can be used as an etch mask for the etching of the bulk material, while the second approach allows for an arbitrary design of the supporting structure.

Figure 2 illustrates the fabrication process. Utilizing silicon substrates with a (110) crystal orientation allows the fabrication of a grid structure with flat sidewalls due to the anisotropic etching behaviour of silicon substrates by potassium hydroxide (Fig. 2a). The substrates are prepared by ultrasonic cleaning in acetone and isopropanol, followed by the native oxide removal in 5 % hydrofluoric acid. The GC material is deposited subsequently with the desired thickness. The slow deposition rate of 1 nm/min results in a precise and reproducible thickness of the window material. The high resilience of the GC material against potassium hydroxide makes it an ideal mask material as it can be easily structured with an oxygen plasma to define the support structure of the final GC transmission window. The bulk silicon is subsequently removed by a wet etch in hot potassium hydroxide, while the silicon grid structure remains wherever the bulk silicon was covered by the patterned GC material. The window is released as the silicon is completely removed and the high compressive stress of the GC material results in wrinkle formation when it is no longer constrained by the substrate.

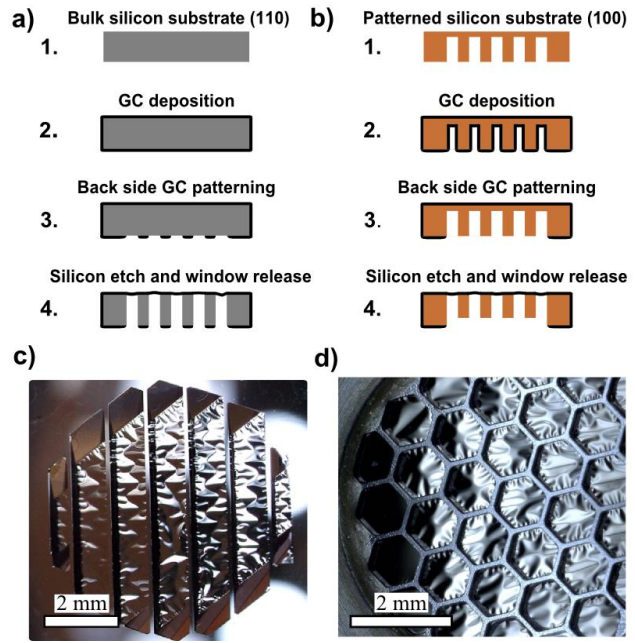

**Figure 2** The process flow for fabricating a bar grid structure with the use of Si (110) substrates is shown in (a) and for a hexagonal support structure in (b). The bottom view images of the resulting GC membranes on the Si support are shown in (c) for the grid bar design and in (d) for the hexagonal structure. The wrinkles form in the free standing GC material due to the compressive stress of the deposited GC material.

Using a pre-structured silicon substrate, prepared by dry-etching, allows arbitrary grid designs that are not limited by the crystallographic nature of the silicon substrate (Fig. 2b).

The used mask layout for bar grid designs, results in a nominal fill factor of 87 %, with a free span width of 1 mm and 160 μm wide supporting Si-bars as shown in Fig. 2c.

The pre-structured substrate is fabricated by etching the major part of the bulk silicon from the backside by a dry etching process, prior to the GC deposition. A hexagonal grid structure with a parallel spacing of 1 mm and a support grid width of 80 μm was chosen and results in a nominal fill factor of 85 % as shown in Fig. 2d.

X-ray transmission measurements were performed as described in detail in references [20] and [28], utilizing a high purity calibration sample and a dual SDD detector setup. This allows the determination of the x-ray transmission of the windows at discrete energy levels corresponding to the energy lines of C $K_\alpha$ (277 eV), Mn $L_\alpha$ (637.4 eV), Cu $L_\alpha$ (929.7 eV), Al $K_\alpha$ (1.49 keV), and Zr $L_\alpha$ (2.04 keV) [5]. Simulations were performed using the data available from Henke et al. [6] to complement the measured, discrete transmission values with continuous data points.

Helium leak testing was performed with a Pfeiffer HLT 570 helium leak tester and values below $1 \times 10^{-10}$ mbar L/s are considered sufficiently leak tight to hold the high vacuum inside the detector housing over the life time of a SDD detector. Leak testing is used to rule out the formation of micro cracks in the window material and in order to monitor the integrity of the GC transmission windows during the various testing scenarios.

Pressure cycle testing, which creates a changing differential pressure across the window, was conducted with a differential pressure of at least 1200 mbar and a frequency of 3 Hz. The maximum and minimum pressure of each pressure cycle was recorded. The low energy transmission windows were cycled at a slower rate of 0.5 Hz and at an atmospheric differential pressure [20].

The optical transmission of the fabricated low energy GC transmission windows was measured with a halogen light source with a spectrum similar to day light with a maximum intensity at a wavelength of 580 nm and a LD didactic compact spectrometer with a specified resolution of 1 nm.

X-ray irradiation was performed with a commercial x-ray tube (Oxford Eclipse 3) with an acceleration voltage of 30 kV, a current of 100 µA, and the GC window was placed at a distance of 4 cm. The resulting count rate totals to $9 \times 10^8$ cps per window area. The total dose was at least $6 \times 10^{12}$ photons which is a 6-fold of the guaranteed irradiation stability of the SDD detector itself.

An ozone containing atmosphere was generated by placing a UV light source emitting 4.5 W at the 184.9 nm wavelength within an airflow that was directed at the GC transmission window.

Bulge tests were performed in order to determine the Young's modulus of the GC window material. The used setup is shown in Fig. 3. A pressurizable sample holder is placed on a motorized stage and a differential pressure used to deflect the freestanding GC film. The resulting membrane deflection was recorded with high accuracy using a Keyence SI-F01 laser displacement sensor with a specified resolution of 1 nm. Profiles of the membrane were obtained by scanning across the membrane geometry at a pre-set differential pressure. The simultaneous measurement of the applied differential pressure was used to estimate the Young's modulus of the window material by relating the strain of the membrane to the pressure induced stress of the GC material.

The membrane dimensions were measured with a calibrated camera, and the film thickness was determined with an atomic force microscopy measurement on a previously patterned region of the thin film.

Raman measurements were obtained with a commercial Raman spectrometer from B&W Tek (inno-Ram-532H) with an excitation wavelength of 532 nm, a specified spot size of 10 µm and an output power of 50 mW. A motorized stage with a pressurizable sample holder allows spatial Raman mapping of the GC window while a defined differential pressure is applied to the GC window. The position of the G peak was determined by using a single Breit-Wigner-Fano (BWF) peak shape for the G peak and a Lorenzian peak shape for the D peak to fit the obtained Raman spectrum, as described by Ferrari et al. [29]. The standard deviation of the extracted G peak position for the same sample was determined to be 0.97 cm$^{-1}$.

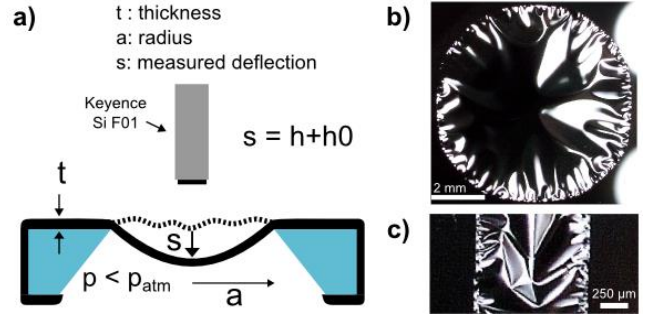

**Figure 3** The setup of the bulge test to determine the Young's modulus of the GC material is shown in (a). A differential pressure is applied to the freestanding GC films and the centre deflection recorded with a displacement sensor. The wrinkles lead to a slack which is accounted for by an additional deflection h0. Round window geometries as depicted in (b) as well as rectangular geometries, shown in (c) are probed.

Finite-element mechanical simulations were performed using Ansys 14.5.

**3 Results and Discussion of Low Energy GC Transmission Windows** The suitability of the windows as a low energy transmission window was evaluated and compared with currently available low energy window solutions. Windows with the bar grid design were fabricated with a GC thickness of 140 nm, whereas the windows with pre-structured hexagonal grid design were fabricated with a GC thickness of 220 nm. Both window types withstand a differential pressure of 1 bar and exhibit a window diameter of 7 mm. The x-ray transmission of the windows was examined by measuring the transmission at discrete energy levels (Fig. 4). In addition, simulations according to the model of Henke et al. [6] are performed with the specific window thickness, assuming a GC density of 2.2 g / cm$^3$.

As shown in Fig. 4, the measured and simulated data are in good agreement. The bar grid design in combination with the reduced GC thickness results in a higher transmission for the energy range of 0.1 keV to 2 keV compared to the hexagonal grid design (Fig. 4a). By comparing the measured transmission values to those of a 1 µm thick GC window without a support grid, an increased transmission below 1 keV is observed. The reduced fill factor of the low energy windows leads to a signal loss for higher energies.

Although the hexagonal grid design exhibits a nominal fill factor of 85 %, the measurements indicate a reduced fill factor of 76 % which is a result of shadowing effects, as the incoming radiation has an angular distribution and not only contains photons normal to the window surface. The fabricated bar grid exhibits a reduced angular dependency as shadowing is only possible in the direction perpendicular to the supporting bars.

Comparing the window transmission of the bar grid design to the transmission of polymer windows [30] in Fig. 4b, shows a superior transmission for the GC window over the whole energy range while the transmission of fluorescent blind silicon nitride windows [11] is similar with an improved transmission below the carbon x-ray line and a reduced transmission between 280 eV and 800 eV.

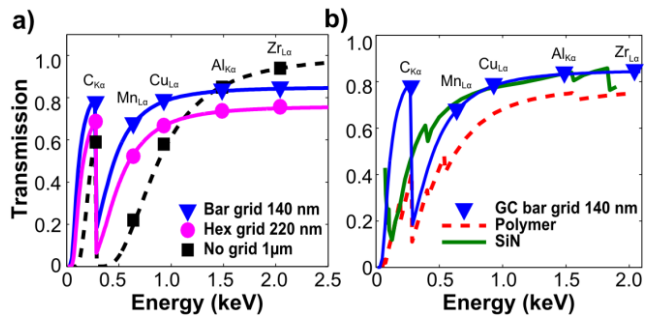

**Figure 4** The measured x-ray transmission for discrete energy lines are displayed in (a) for GC windows with a bar and a hexagonal grid design. They are compared to a 1 µm thick GC window without a support grid [20]. The continuous data points are simulated according to the model of Henke et al. [6]. In (b), the transmission of a bar grid window design with a GC thickness of 140 nm is compared to the transmission of polymer windows [30] and fluorescent blind silicon nitride windows [11].

Both GC window types were pressure cycle tested with an atmospheric differential pressure for 17 k cycles. They did not show detectable visible degradation nor did the cycle test show an increased helium diffusion, as can be seen in Fig. 5a. The high gas tightness of the GC material allows the vacuum encapsulation of the detector modules which has not been possible with polymer windows.

Figure 5b shows the optical transmission measured for the bar grid layout with a GC thickness of 140 nm. The optical attenuation of the GC window is higher than the values for polymer windows that incorporate an additional aluminium light blocking layer. No comparison data is available for silicon nitride transmission windows. The high optical attenuation of GC avoids the otherwise needed light blocking layer. Therefore, delamination or cracking of this additional layer during the life time of the detector can be ruled out for GC windows.

The designed and fabricated GC windows fulfil the requirements for low energy x-ray transmission windows. The bar grid window design with a GC thickness of 140 nm can thus replace polymer windows, offering the advantages of a higher x-ray transmission, gas tightness and an increased optical attenuation while exhibiting high resilience towards cyclic stress.

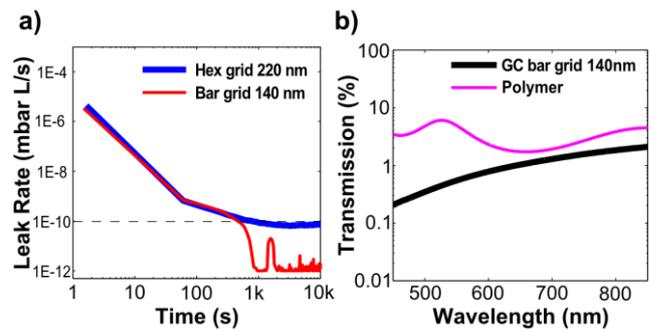

**Figure 5** The helium leak rate for both window designs is shown in (a) after each window was subjected to 17 k pressure cycles of an atmospheric differential pressure. The optical transmission of a bar grid GC window with a GC thickness of 140 nm has been measured in (b) and compared to the measured values of a polymer window.

**4 Results and Discussion of the GC Material Properties** GC transmission windows without a support grid, an open diameter of 7 mm and a GC thickness of 1 µm were tested for radiation hardness and against ozone exposure. The GC windows were exposed to a total ionizing dose of $6 \times 10^{12}$ photons. No visible deterioration of the GC surface or increased helium diffusion (Fig. 6a) could be observed after exposure. In a second experiment, GC windows were placed in an ozone containing environment for a total of 3 days for ozone exposure testing. Subsequent helium leak tests did not result in an increased helium diffusion compared to a test performed prior to the ozone treatment (Fig. 6b).

A thorough understanding of the material properties of GC is necessary for further optimization of the window design. An important material property describing the behaviour of the material is the Young's modulus of the GC material.

A common method to determine the Young's modulus of a thin film material is the bulge test [31]. This test uses the deflection of a thin film due to an applied differential pressure and evaluates the Young's modulus based on the assumption of a homogeneous stress distribution and a spherical bulge shape [32].

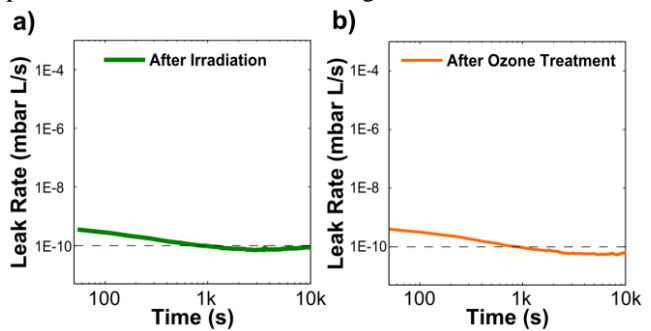

**Figure 6** The helium leak rates for 1 µm thick GC transmission windows are plotted after the window was subjected to ionizing radiation (a) and an ozone containing atmosphere (b). The helium leak tests did not show increased helium diffusion.

A model describing the centre deflection of a round membrane was developed in 1958 by Beams et al. [33] under the assumption of homogenous biaxial stress and strain across a spherical bulge, allowing a simple extraction of the biaxial modulus.

The simplified model ignores the uniaxial stress and strain state that occurs at the anchoring points of the deflected membrane and the deviation of the bulge shape from a spherical shape.

This has led to the improvement of the deflection model by various groups using energy minimization [32] and finite-element simulations [34], modifying the model in order to better represent the actual bulge state. Vlassak et al. [35] postulated the advantage of bulging long rectangular geometries in order to avoid the transition of a biaxial stress and strain state at the membrane centre to a uniaxial stress and strain state at the anchoring points. Rectangular geometries with a high aspect ratio exhibit a solely uniaxial stress and strain state across the narrow region of the bulge shape [36]. The presented models assume free standing films that are stress free or exhibit tensile stress. This is in not the case for the GC windows as the large compressive stress of the GC material leads to the observed wrinkle formation.

Figure 7a shows the results of the bulge test experiments on a 7 mm wide window. The centre deflection is extracted from line scans across the window geometry. The high compressive stress of the deposited GC material leads to a large slack of the relaxed free standing GC film. Therefore, a small differential pressure of 6 mbar already results in a large deflection of 157 µm in the centre. The large slack of the film, in combination with the very smooth and highly reflective surface, results in the signal loss of the laser sensor while it scans across the steep window slopes near the anchoring points of the window, as well as during the scan of the unloaded window due to the present wrinkles. In the following bulge tests, a differential pressure of 6 mbar is used as a preload step to smoothen the wrinkles in order to avoid the slack window state. The data shown in Fig. 7b were obtained from 4 samples, three of which are circular windows with an open diameter of 6.9 mm (#1), 7.3 mm (#2), and 5.4 mm (#3) and a thickness of 950 nm (#1), 1100 nm (#2), and 1000 nm (#3) respectively. The other sample has a rectangular membrane geometry with an aspect ratio of more than 4 which is seen as sufficient to assume a uniaxial strain distribution across the narrow region [36], with a film thickness of 140 nm (#4) and a span width of 1.04 mm (#4).

The obtained data was subsequently fitted (not shown) to the bulge test models for spherical and rectangular membranes, but the results were poor and inconsistent. This is assumed to be due to the fact that the models were designed to comply with the shape, stress and strain states of a stress free or taut thin film, which is not the case for the probed GC membranes [32].

Finite-element mechanical simulations were performed in order to provide a better understanding of the stress and strain state of a thin film with an intrinsic compressive stress. Figure 7c shows the results of the simulation of a circular membrane with a diameter of 1 mm, a film thickness of 1 µm and an initial compressive stress of the thin film of 400 MPa. For the simulation an isotropic Young's modulus of 130 GPa, a Poisson's ratio of 0.16, a material density of 2.2 g / $cm^3$ [20] and a load of 1 bar differential pressure were chosen. The used Poisson's ratio being the value of bulk graphite in the basal plane, which has also been suggested for graphene [37, 38] and is therefore used as an estimate for the GC material.

The resulting strain intensity, plotted in Fig. 7c, not only shows increased strain values near the anchor points but also irregularities further towards the centre of the thin film which appear in a periodic manner. Such behaviour is not apparent in deflected membranes without intrinsic compressive stress. The assumption of biaxial strain is validated only for the central region of the window while a large uniaxial tensile strain occurs at the anchoring points. The simulations show that circumferential compressive stress is still present near the anchor points, even at a high differential pressure load, which was also identified by Small et al. [32] and seen as a major obstacle for bulge testing of films exhibiting intrinsic compressive stress.

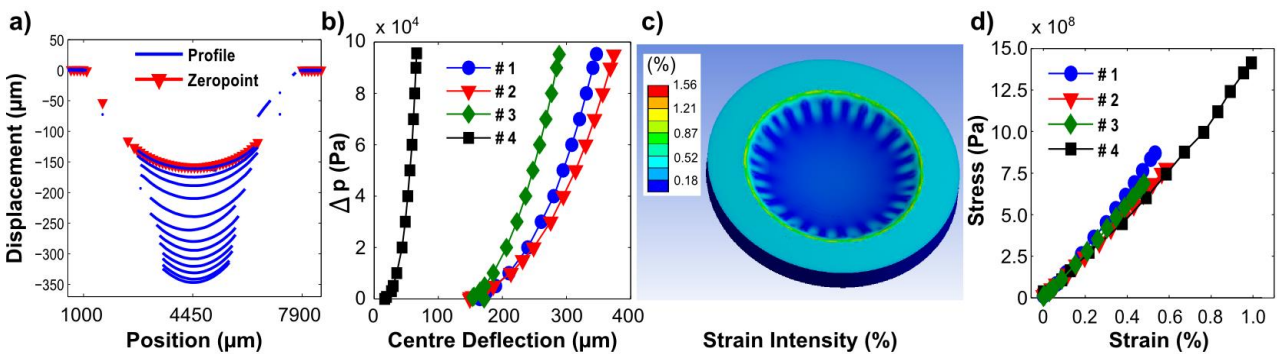

**Figure 7** The centre deflection of the GC window was measured by scanning across the window. The deflection profile corresponding to a pre-set pressure, which was increased from 0 to 950 mbar in a step wise manner is shown in (a). The compressive stress of the GC material leads to a large slack and a differential pressure of 6 mbar is used as a preload step to reduce the impact of the present wrinkles. Pressure deflection curves are shown for 4 samples in (b). The centre deflection depends on the GC thickness (#1: 950 nm, #2: 1100 nm, #3: 1000 nm, #4: 140 nm) and opening diameter (#1: 6.9 mm, #2: 7.3 mm, #3: 5.4 mm, #4: 1.04 mm). Finite element simulations show a large increase of the strain values and a resulting compressive stress state near the anchor points of the window, which hinders fitting the deflection curves to available models (c). Stress–strain curves are therefore used to determine the Young's modulus of the GC material (d). The low thickness of #4 leads to a large stress value at the maximum applied differential pressure compared to the other samples.

In general, the biaxial modulus $Y$ describes the relationship between the biaxial stress $\sigma$ and the biaxial strain $\varepsilon$ of the material under test, as shown in Eq. 1.

$$Y = \frac{\Delta\sigma}{\Delta\varepsilon} \quad (1)$$

Instead of fitting the obtained pressure deflection curves to the available models, we extracted the stress and strain of the membrane, assuming a homogeneous stress and strain distribution across a spherical bulge shape. Stress-strain curves are then used to extract the Young's modulus. The strain $\varepsilon$ is calculated in Eq. 2 by the relative length change [32] induced by the load, where b is the measured arch length and $b_0$ the arch length at a differential pressure of 6 mbar.

$$\varepsilon = \frac{b - b_0}{b_0} \quad (2)$$

Geometric considerations of the arch length of a spherical bulge shape, as shown in Fig. 3a, result in Eq. 3, where s is the centre deflection, a the radius of the window, and $b_0$ the arch length of the bulge shape at a differential pressure of 6 mbar and t the film thickness.

$$\varepsilon = \frac{\frac{(s+a^2)}{s}\arcsin\left(\frac{2as}{a^2+s^2}\right) - b_0}{b_0} \quad (3)$$

Equation 4 describes the homogeneous stress distribution of a thin walled pressure vessel [32] which is used to estimate the stress of the GC material, where P is the measured differential pressure.

$$\sigma = \frac{P(s^2 + a^2)}{4st} \quad (4)$$

Being aware of the implications that arise from the intrinsic compressive stress we use the stress-strain curves (Fig. 7d) to give an estimate of the Young's modulus of the GC material. Small et al. [32] introduced Eq. 5 to derive the Young's modulus $E$ from the biaxial modulus by taking the transition from biaxial strain in the bulge centre to uniaxial strain at the anchoring points into account, where $\nu$ is the Poisson's ratio of the GC material.

$$E = \frac{(1-\nu)}{(1-0.241\nu)} Y \quad (5)$$

Evaluation of the measured data results in an estimated Young's modulus of 142 GPa for #1, 115 GPa for #2, 124 GPa for #3 and 136 GPa for #4. It should be noted that the rectangular sample (#4) experiences an increased stress by a factor of 2 compared to Eq. 4 due to the assumption of a cylindrical bulge shape The Young's modulus is extracted from the slope of the stress-strain curve using Eq. 6 due to the uniaxial strain state of the thin film [36].

$$E = Y(1-\nu^2) \quad (6)$$

The calculated stress for #4 reaches a maximum value of 1.4 GPa at a differential pressure of 950 mbar, which is much higher than the values of the other samples (Fig. 7(d)). This is due to the reduced thickness of only 140 nm while the identical differential pressure was applied. The obtained data suggests that the Young's modulus is largely independent of the GC thickness.

Tensile testing of materials is used to determine the ultimate tensile strength of a material which describes the stress required to induce material failure [39]. Xiang et al. [40] used the bulge test experiment to estimate the ultimate tensile stress of thin films by determining the stress value at the maximum differential pressure reached before the film failed. The high tensile strength of the GC window material is apparent if we compare the maximum stress value of #4 (1.4 GPa), which was still intact at this differential pressure, to literature values of Be which reaches up to 454 MPa [41]. It should be noted that the calculated stress value should be seen as a lower minimum as the finite element simulations indicate a significantly increased stress value near the anchor points (Fig. 8a) of the window compared to the calculated stress value used for the stress-strain curves (Eq. 4).

This discrepancy becomes evident if we use the output, namely the centre deflection, of the simulation and use

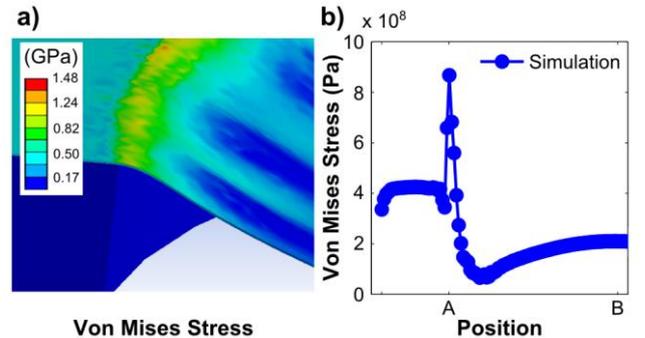

**Figure 8** Finite element simulations indicate a high stress near the anchor points of the window (a). The large increase at position A is found near the anchor points of the free standing film and is approximately 4-fold of the stress experienced at the centre of the window at point B (b).

Eq. 4 to calculate the stress. The obtained value of 248 MPa is much lower than the plotted (Fig. 8b) maximum stress of 933 MPa at point A which is at the anchor point of the membrane. The ultimate tensile strength of the material is therefore tested exactly at the edge of the free standing film where there is an increased stress compared to the rest of the window.

The results of the simulation indicate that the compressive stress of the GC material and the resulting, complicated strain state of the GC window under a differential load limits the validity of the obtained values for the Young's modulus as the systematic error is not known.

Therefore, for a better understanding, Raman spectroscopy is employed in order to probe the local strain state of the $sp^2$-bonded carbon of the GC window. Taylor et al. [42] were able to determine the stress and strain in a pyrolytic carbon by evaluating the position of the G peak. Tensile stress stretches the carbon bonds and leads to phonon softening, shifting the G peak to lower wavenumbers [43, 44]. Raman spectrums were recorded along the indicated line shown in Fig. 9a of a GC window with and without an applied differential pressure. Figure 9b shows the position of the G peak depending on the position on the transmission window. Raman measurements obtained on tilted GC films did not result in a shift of the G peak position ruling out a dominant influence of the incident angle. A large upward shift of 4 cm$^{-1}$ is observed in a Raman spectrum that is recorded in a region where the GC material is still constrained by the silicon substrate compared to a spectrum obtained on the relaxed, free standing GC film. On the other hand, applying a differential pressure load of 900 mbar leads to a large tensile strain and results in phonon softening, shifting the G peak of a spectrum obtained in the centre of the GC window by 5 cm$^{-1}$ to lower wavenumbers compared to the spectrum obtained without the applied differential pressure (Fig. 9b).

Raman spectrums of the relaxed and deflected GC window indicate high compressive strain of the GC material that is still constrained by the silicon frame as seen by the G peak position being at higher wavenumbers (1594 cm$^{-1}$) compared to the G peak position of the relaxed GC film at 1590 cm$^{-1}$. The G peak position of the deflected window region is not uniform, with a maximum shift to lower wavenumbers at the centre of the window, which indicates a non-uniform strain distribution. This is contrary to the assumptions made during the bulge test, where a homogeneous strain distribution was assumed. Mohiuddin et al. [45] demonstrated that uniaxial strain in graphene leads to a reduced shift of the G peak position compared to the shift induced by biaxial strain. The reduced shift of the G peak near the anchoring points and the gradual change to higher wavenumbers, as we move from the deflected centre towards the edge of the free standing GC film, could thus also be a result of the transition from a biaxial strain state at the centre to a uniaxial strain state at the edge of the free standing film [38].

In contrast to the simulations, the strain enlargement at the anchoring points was not observed by the Raman measurements. This is attributed to the relatively large spot size of the used Raman microscope.

Figure 9c shows the relationship between the applied differential pressure and the Raman shift of the G peak position. These spectra were obtained at the centre position of a circular membrane where biaxial strain is present, according to theory and our finite-element simulations. [32] The G peak position is correlated to the strain of the GC material that was obtained by the previous bulge test experiment. Fitting the Raman shift of the G peak to the calculated strain (Fig. 9d), results in a linear relationship with a shift of the G peak position of -9.4 cm$^{-1}$ per 1 % of strain. The value $\partial \omega_G / \partial \varepsilon$ is used to quantify the slope of the linear fit, $\partial \omega_G$ being the relative change of the position of the G peak and $\partial \varepsilon$ the corresponding change of the strain.

The obtained value fits very well to $\partial \omega_G / \partial \varepsilon$ = -9.5 cm$^{-1}$ / % which was determined by Taylor et al. [42] by indentation experiments on pyrolytic carbon that was deposited on a quartz fibre, which is surprising considering the implications of the compressive intrinsic stress.

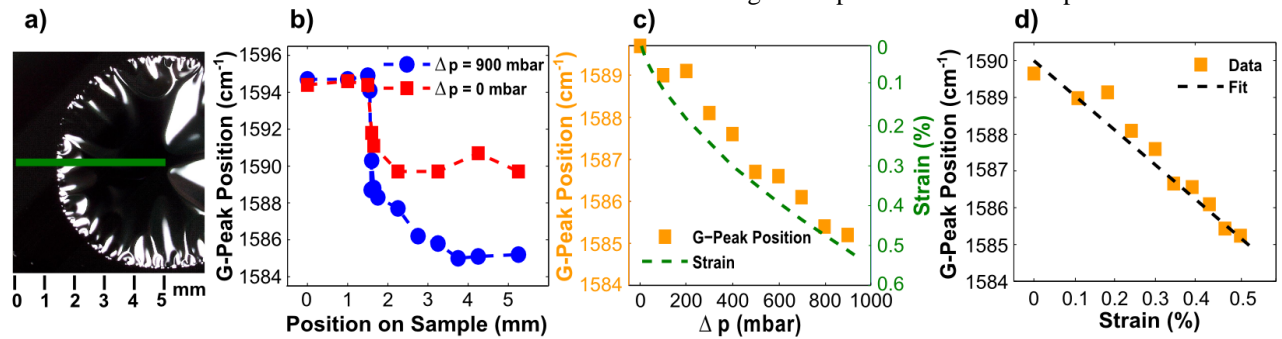

**Figure 9** The top view image of a 7 mm wide, 1 μm thick GC window in (a) indicates (red line) where Raman measurements were performed. The resulting position of the G peaks along this path are plotted in (b). An applied differential pressure shifts the G peak position towards lower wavenumbers. An inhomogeneous shift is apparent as well as a large compressive strain that leads to a blue shift of the G peak position in the region where the GC material is still in contact with the silicon substrate (b). The G peak position is plotted versus the applied differential pressure in (c) for spectrums obtained at the centre of the window. The shift of the G peak position can be related to the calculated strain values that were obtained by the bulge test experiments. This results in a linear dependency of the G peak position on the experienced strain, displayed in (d).

The results of the Raman mapping and the extracted $\partial\omega_G / \partial\varepsilon$ of the strain induced G peak shift seem to indicate that the estimation of the strain is better than expected.

While the estimated Young's modulus of the GC material is substantially higher than the literature values of the Young's modulus for pyrolytic carbon (10 - 45 GPa) [46], the obtained value $\partial\omega_G / \partial\varepsilon$ is similar. An enormous difference is apparent if this is compared to the $\partial\omega_G / \partial\varepsilon$ value of graphene. Metten et al. [38] used a monolayer graphene blister with a diameter of 4 μm and a bulge test setup to observe a change of $\partial\omega_G / \partial\varepsilon = -57 \pm 5$ cm$^{-1}$ / %. And Mohiuddin et al. [45] reported a similar value of $\partial\omega_G / \partial\varepsilon = -63$ cm$^{-1}$ / % for graphene under biaxial strain.

We can use the extracted value of $\partial\omega_G / \partial\varepsilon = -9.4$ cm$^{-1}$ / % and the measured G peak shift of 4 cm$^{-1}$, between the G peak position of substrate constrained and relaxed GC material, to estimate the compressive stress of the GC film still bound to the silicon substrate. This results in an estimated compressive stress of the deposited GC material of approximately 500 MPa. The abrupt transition of the compressive strain state to a tensile strain state of the GC material at the anchoring points of the window, as well as the high strain levels in this region, indicated by the finite-element simulations, make the observed high resilience [20] of the GC windows against cyclic loading even more remarkable and demonstrates the high mechanical stability of the GC material.

A 1 μm thick GC transmission window was therefore pressure cycle tested at an increased differential pressure of at least 1200 mbar. The window did not show signs of degradation even after 10 million pressure cycles (Fig. 10a). A helium leak test performed after this cycle test did not reveal an increased helium diffusion (Fig. 10b). Raman measurements obtained before, during and after the cycle experiments did not resolve reordering of the bonding structure of the GC – at least within the resolution of our equipment.

**4 Conclusion** It was demonstrated that the energy range of GC x-ray transmission windows can be further extended to the low energy range by choosing an appropriate GC film thickness and support grid. The presented bar grid window design could be improved even more, in the future, by optimizing span and support grid width. This would allow for a further reduction of the GC thickness and an even higher transmission. Further investigation will be required to understand in detail what makes this material so durable, especially as the maximal stress occurs in the location where the stress changes abruptly from compressive to tensile stress as shown by Raman measurements.

**Acknowledgements** This work has been financially supported by the Bavarian Ministry of Economic Affairs and Media Energy and Technology under contract number MST-1210-0006/ BAY 177/001. The authors would like to thank Silke Boche and Lukas Holzbaur for their ongoing contribution and support.

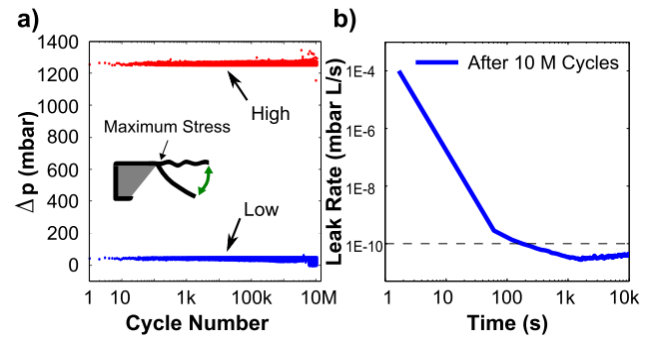

**Figure 10** The result of a pressure cycle test with a 1 μm thick GC window with an open diameter of 6.9 mm is shown in (a). The GC window did not show signs of degradation and was still intact after 10 million pressure cycles. A subsequent helium leak test did not show increased helium diffusion as shown in (b).